\documentclass{ifacconf}
\usepackage{graphicx}      
\usepackage{epsfig}
\usepackage{subcaption}
\usepackage{natbib}        
\usepackage{amsmath}
\begin{document}
\begin{frontmatter}

\title{Characterizing pitch and roll torque coupling in insect-sized flapping-wing robots using a microfabricated gimbal} 

\thanks[footnoteinfo]{This work was partly supported by the NSF award Nos. 2319060, 2054850, and 2235207.}
\thanks[video]{A supplemental video for this paper is available at: https://youtu.be/COBHHIgYGRU}

\author[First]{Aaron R. Weber} 
\author[Second]{Daksh Dhingra}
\author[Third]{Sawyer B. Fuller} 

\address[First]{Department of Mechanical Engineering, University of Washington, Seattle, USA (e-mail: aweber6@uw.edu).}
\address[Second]{Humanoid, Boston, USA (e-mail: dakshdhingra93@gmail.com)}
\address[Third]{Department of Mechanical Engineering and Paul G. Allen School of Computer Science, University of Washington, Seattle, USA (e-mail: minster@uw.edu)}

\begin{abstract}
Sub-gram flapping-wing flying insect robots (FIRs) are challenging to model because of mechanical complexity in their wings, unsteady aerodynamic flow, and the difficulty of making precise measurements at small scale. Coupling effects between roll and pitch torque actuation have not previously been measured because a two-axis sensor that is sensitive enough has not been realized. To address this shortcoming, we introduce a microfabricated gimbal design capable of precisely and simultaneously measuring roll and pitch torques as well as thrust. We then used it to measure the extent to which a pitch torque command affects roll torque and vice versa on a 180~mg piezo-actuated flapping-wing flying platform. Our results show a high coefficient of determination in the linear regression for both pitch (0.95) and roll (0.98) and low cross-correlation coefficients (–0.001 and –0.085, respectively) across the full range of simultaneous torque commands, indicating negligible cross-axis coupling. Similarly, thrust force deviates by a maximum of only 5.8\% from the mean thrust value. These results validate the assumption that pitch and roll can be considered independently in control and will inform future models of how inputs affect the aerodynamics of resonant flapping-wing systems.

\end{abstract}


\end{frontmatter}

\section{Introduction}\label{sec:introduction}

\begin{figure}[tbp]
    \includegraphics[width=\columnwidth]{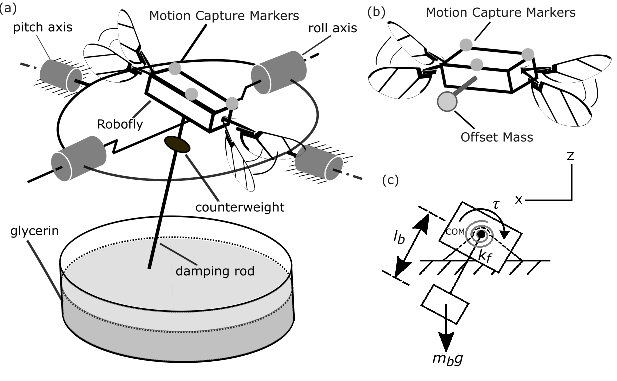}
    \caption{Illustration of the flexured-gimbal device principle and operation. (a) Diagram of the principle of the flexured-gimbal force/torque sensor. The flexures are positioned such that the roll and pitch axes of the device both intersect with the approximate center of mass of a Flying Insect Robot (FIR) such as the UW Robofly. A rod immersed in glycerin provides damping to eliminate undesirable resonant oscillations. The gimbal is supported by a lightweight structure (not shown) that is placed on a sensitive balance to measure thrust force. (b) For validation, a known torque is applied to the FIR by attaching a known offset mass at a known distance from the center of mass (COM). Compensatory free-flight trim values are compared to the torque measurement from the gimbal. (c) Simplified model of flexured-gimbal torque sensor under static conditions. An FIR at location COM coincident with a spring-like flexure joint rotates a small amount due to a torque $\tau$ applied by the wings. This is counteracted by spring-like flexure stiffness $k_f$ and a pendulum counterweight $m_b$.}
    \label{axisFig}
\end{figure}

Flying insect-sized robots (FIRs) weighing less than a gram have to date used flapping wings inspired by insects to propel themselves. Because they are small and low weight, FIRs are expected to be better at reaching tighter spaces than is possible with more common four-rotor drones, and may be able to power themselves with sunlight due to favorable scaling physics~\citep{elkunchwar,jafferisnature2019}. Due to these properties, they have promising potential in applications like search and rescue, inspections and inventory in manufacturing plants, and detecting gas leaks. 

At insect scale electromagnetic motors become inefficient, requiring other means of actuation for robots. Inspired by biology, many flapping-wing FIRs use piezoelectric actuators connected to the wings through a transmission system~\citep{perez2011first,fuller2019four,beeplus}. Electrostatic actuators such as piezoelectric cantilevers can operate at high efficiency even at the centimeter scale, unlike electromagnetic motors, in which friction forces and heat dissipation in coils increasingly dominate as the size gets smaller~\citep{Trimmer1989}. 

Moving from rotors to piezoelectric actuators with flapping wings provides the efficiency needed for FIRs, but the unsteady flow phenomena of flapping wings, such as leading edge vortices, delayed stall, and spanwise flow, greatly increase the difficulty of modeling. Previous work in drone modeling includes~\citet{Mahonyquad2012}, which introduced a motor model that converts the input PWM signal to the rotor speed for their quadrotor systems. \citet{KarasekDelfly2018} used an electronic speed control in conjunction with an angular rate sensor to achieve precise motor speed control on a 28~g flapping-wing robot. Modeling and control of much larger ($>$100~g) flapping-wing bird robots has also been demonstrated \citep{Gerdes2014,Zufferey2021,Nekoo2023}. These models rely on the well-known and consistent properties of motors. In our experience, obtaining feedback about intermediate states or a reliable model like these for sub-gram FIRs is challenging because of their inherent variability and the unsteady flow phenomena involved in flapping-wing flight. The performance of the piezoelectric actuator varies between devices, and the flapping process places a great deal of stress on its flexure joints, resulting in a short device lifespan \citep{malka2014principles}. This frequently causes the FIR dynamics to change during experimentation.

Control of FIRs~\citep{chirarattananon2013adaptive,fuller2019four,ma2013controlled} has to date depended on robust feedback control systems to compensate for modeling uncertainties, particularly in how torque commands map to output torques and thrust. However, in practice, approaching control in this way increases the amount of trial and error needed to tune the gains of a controller because it is harder to establish the source of undesirable performance. For example, oscillations might be attributable either to an inaccurate torque mapping model or to suboptimal gains. 

Previous work characterizing FIRs investigated only a single axis at a time. For example, ~\citet{perez2011first} found that thrust is approximately linear with driving voltage on a piezo-actuated FIR, and~\citet{ma2012design} found that the same holds for each of the three perpendicular torque axes. \citet{Wu2026} utilizes a gimbal apparatus similar to our work to tune pitch and roll controls in an FIR, though two single-axis gimbals are used rather than a single two-axis gimbal and the hummingbird-inspired FIR used is much larger than the insect-inspired FIRs this work targets. The goal of our work is to further improve the characterization of FIRs by developing a method to precisely measure \textit{multi-axis effects} in FIRs. In particular, it is unknown how voltage commands on a given axis affect its other degrees of freedom. This knowledge will help improve aerodynamic model fidelity, flight controller performance, and the evaluation of new designs.

Measuring torques at the scale of FIRs is challenging because the small torques involved preclude using off-the-shelf sensor hardware. The smallest commercially available multi-axis torque sensor, the ATI Nano17 Titanium, has a resolution of 8~$\mu$Nm. This is more than an order of magnitude higher than what is needed to accurately measure FIR torques. It has been demonstrated that strain gauges with proper filtering can be used to measure micro-forces in \citep{HADDAB2009457}, but the noise amplitude is still more than an order of magnitude higher than torques output by FIRs \citep{ma2012design,firTorques}. Previously, capacitance-based sensors have been introduced that have sufficient resolution to measure torque about a single axis on an FIR~\citep{finio2011torques} and force along two simultaneous axes~\citep{Wood_2009}. Capacitive sensors, however, require expensive specialized hardware that costs \$1000 or more. Furthermore, a two-axis version capable of measuring two torque axes at once has not yet been demonstrated at the scale of an FIR. Force sensors utilizing probes with the required resolution for FIR measurements have also been designed \citep{MAROUFI2018198}, but these sensors only operate on a single axis.

Here, we introduce an alternative torque-sensing system (Fig.~\ref{axisFig}a) that does not require pricey capacitive hardware and instead uses a microfabricated gimbal and camera-based motion capture system that is commonly available in a robotics laboratory, or potentially an inexpensive inclinometer ($<$~\$10). By doing so, we were able to perform the first simultaneous two- and three-axis torque and force measurements on an FIR platform. We then used this sensor to characterize cross-axis coupling, a measurement that would have been much more difficult on a single-axis sensor given the limited flight lifetime of current FIR designs.

\section{Principle of Operation}
\begin{figure}[tbp]
    \includegraphics[width=.9\columnwidth]{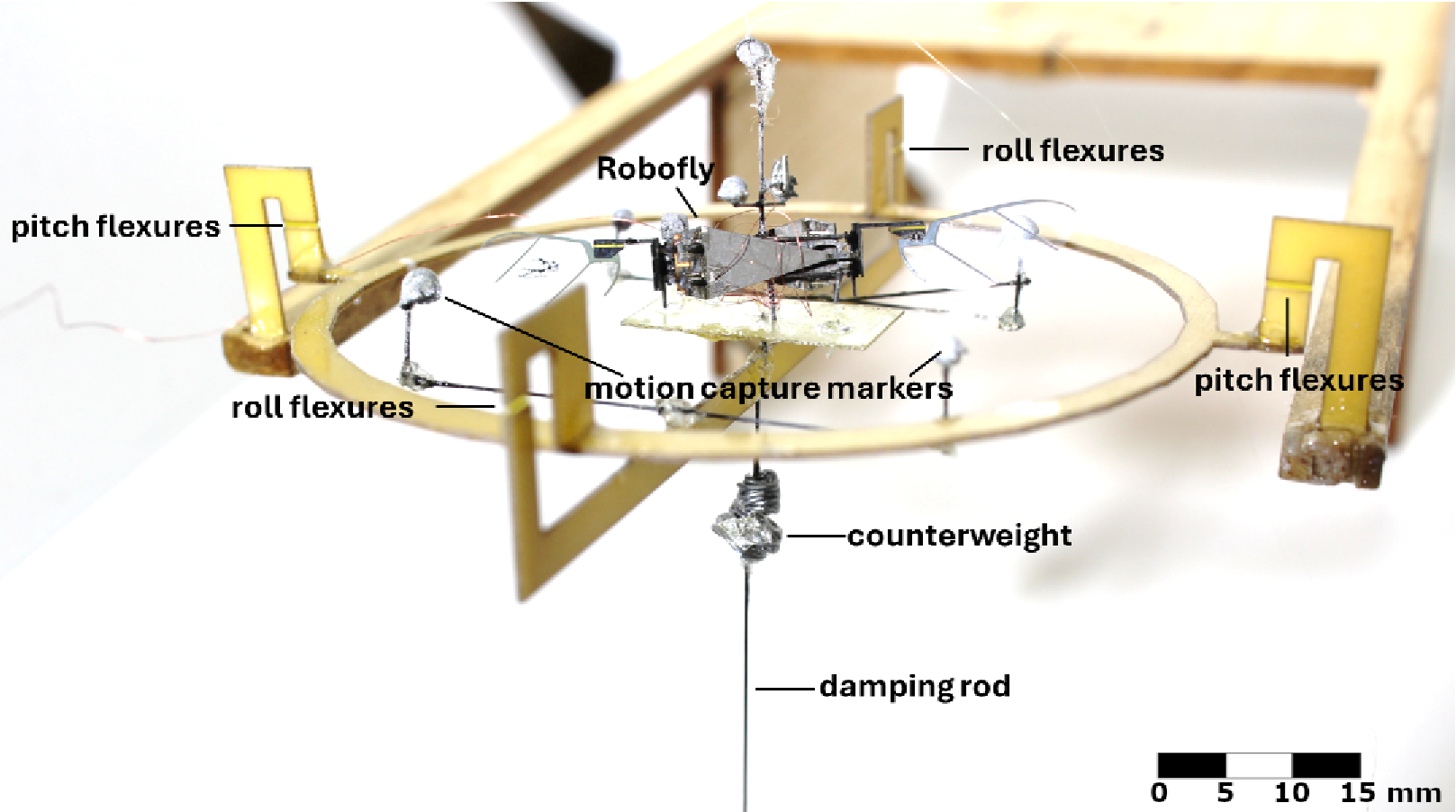}
    \caption{Photograph of the flexured-gimbal device with an FIR attached. Not shown in the figure is the glycerin petri dish below. Readout is accomplished in this case using a camera-based motion capture system. The structure supporting the gimbal (seen in the distance) is placed on a sensitive balance (scale) to measure thrust force. }
    \label{deviceLabels}
\end{figure}

Like capacitive and strain gauge systems, our system reacts to a force or torque by moving a small amount. However, the spring-like restoring torque in our system is lower, so that the angular deflections are larger. This allows the deflection to be measured using less precise hardware such as a camera-based motion capture system or an accelerometer (Fig.~\ref{axisFig}a). This approach reduces the bandwidth of our sensor to below 1~Hz, but that is sufficient to measure torques on a stroke-averaged basis, which is our primary interest here. 

To perform the desired measurements, we introduce a system that is conceptually similar to the two-axis device introduced in~\citet{ddhingraTrimming}. In both systems, a microfabricated system constrains the robot to rotate around two axes---pitch and roll---while keeping all other degrees of freedom fixed. The device is attached to ground through microfabricated flexure joints. Each axis is subject to a spring-like linear restoring torque that causes the robot to remain upright at equilibrium. Both devices additionally aim to minimize the effect that thrust forces have on the measured torque output by placing the center of mass (COM) of the robot so that the two perpendicular gimbal rotation axes intersect the thrust vector of the flapping wing system. In both systems, this is done  by incorporating a small flat surface, termed the ``table,'' on the gimbal, positioned such that an FIR placed on it will be level with the gimbal flexures. Before the FIR is added, the zero-torque position of the gimbal is measured, and then the FIR is placed and then translated in small increments by hand along its pitch and roll axes until the gimbal returns to the zero position. At that point the FIR is bonded to the table with cyanoacrylate (CA) glue. Thin carbon fiber rod ``legs'' attached to the FIR provide a compliant bonding location that can bonded to and removed from the table without putting stress on the frame or actuation components of the FIR. For later free flights, the FIR is de-bonded by melting the glue using a soldering iron. 

The new device, shown in Fig.~\ref{deviceLabels}, incorporates a number of improvements over the previous design introduced in~\citet{ddhingraTrimming}:
\begin{enumerate}
    \item The new design incorporates a ring (a ``flexured gimbal'') large enough not to impede flow from the FIR's wings. This modification allows the flexure axes to intersect the approximate \textit{center of mass of the vehicle}, rather than simply intersecting its approximate \textit{thrust vector} as in~\citet{ddhingraTrimming}. Doing so ensures that if the flapping wings produce a non-ideal thrust force, that is, one that has either a longitudinal (along the roll axis) or lateral (along the pitch axis) component in addition to its desired vertical (yaw axis) component, that force will not result in a confounding torque on the gimbal's flexures. 
    \item A damper is added to reduce unwanted oscillations
    \item A precision scale is added to measure thrust force
    \item Instead of just finding zero-point torque (i.e. trim value), the new system's flexure joint stiffnesses are precisely calibrated so that torque values can be measured. We validated our calibration using free-flight experiments with fixed torques that were known ahead of time. 
\end{enumerate} 

\subsection{Torque Measurement About Flexure Axes }
The dynamics of the gimbal system (shown in  Fig.~\ref{axisFig}c) subject to a torque $\tau$ applied by the flapping wings is
\begin{align}
    \begin{aligned}
       \tau &= m_b g l_b \sin\theta - m_R g l_R \sin\theta + k_f \theta + b\dot\theta + I\Ddot{\theta}
    \end{aligned}\label{eq:dynamics}
\end{align}  
where $m_R$ and $m_b$ are the masses of the robot and counterweight, $l_R$ and $l_b$ are the mounting distance from the axis of rotation for the robot and the counterweight, $b$ is a coefficient approximating the damping of the rod (which is thin and of negligible mass)  in the glycerin dish, $k_f$ is the flexure stiffness of the device, and $I$ is the moment of inertia of the system about a given axis.

In the new version of the device introduced here, the axis of rotation intersects the approximate COM so that $l_R\approx0$, meaning that the $m_R l_R \sin\theta$ term can be neglected. For our application, the measurements are taken when the robot is at steady state, and as such the $\dot\theta$ and $\Ddot{\theta}$ terms will go to zero. Under these conditions and using the small-angle approximation, Eq.~\eqref{eq:dynamics} reduces to 
\begin{align}
    \begin{aligned}
       \tau = k_s \theta,  
    \end{aligned}
    \label{equation_torque_estimate}
\end{align}
where 
\begin{align}
    \begin{aligned}
       k_s = \frac{\tau}{\theta} = m_b g l_b + k_f   
    \end{aligned}
    \label{eq:sensitivity}
\end{align}
is the total stiffness of the device. The sensitivity of the device is inversely proportional to the device stiffness, and can be increased by reducing the mass of the counterweight $m_b$. However, making the sensitivity too high could result in deflections that enter the non-linear zone of the elastic flexure joint.

\subsection{Force Measurement}\label{sec:force}
The gimbal is supported by a mechanical structure placed on precision balance. The structure positions the gimbal and FIR to the side of the scale so that its plate is not subject to aerodynamic forces; the petri dish below the gimbal is placed far enough below to avoid the aerodynamic ground effect. By placing various masses either at the center of the balance's plate or at the laterally-displaced position of the FIR, we were able to confirm that the position at which a force is applied has essentially no effect on the force measurement. 

\subsection{Actuating Torques}
The input to each piezoelectric actuator (one for each wing) on the UW Robofly is a sinusoidal signal: 
\begin{align}
    V_1(t) &=  \frac{1}{2}V_b - V_o + \frac{(A + \delta A)}{2}\sin(2\pi f t) \label{v1_eq} \\
    V_2(t) &= \frac{1}{2}V_b + V_o - \frac{(A - \delta A)}{2}\sin(2\pi f t) \label{v2_eq}
\end{align} 

where $A$ is the baseline peak-to-peak amplitude, $f$ is the flapping frequency, and $V_b$ is the bias voltage for the piezoelectric cantilevers (250~V for all experiments in this work). There is a negative sign in front of the sine term in $V_2$  so that that the wings move in tandem forward and backward at the same time despite having identical piezo-transmission-wing actuation units that are oriented in opposition \citep{ChukewadTRO}. When a high-voltage power system is incorporated, this will also allow energy from one actuator to pass to the other for greater efficiency~\citep{James2021}.
Thrust from each wing is approximately proportional to the amplitude $V_{amp}=\frac{A\pm\delta A}{2}$ of the driving signal.  A positive roll torque is produced by  increasing the amplitude of the left wing relative to the right wing, so that the difference in amplitudes is equal to $ \delta A= V_{amp1}-V_{amp2}$. A positive pitch torque is produced by altering the position of the stroke-averaged center of aerodynamic thrust~\citep{ddhingraTrimming} forward of the CM. This is achieved by changing the mean of the sinusoidal signal by an offset voltage, $V_o$. Note that this convention removes an erraneous factor of 1/2 that appeared in front of $V_o$ in~\citet{dhingra2025modeling}. 

\section{Experimental Apparatus}
\begin{figure}[tbp]
\centering
\begin{subfigure}{\linewidth}
  \centering
  \includegraphics[width=\columnwidth]{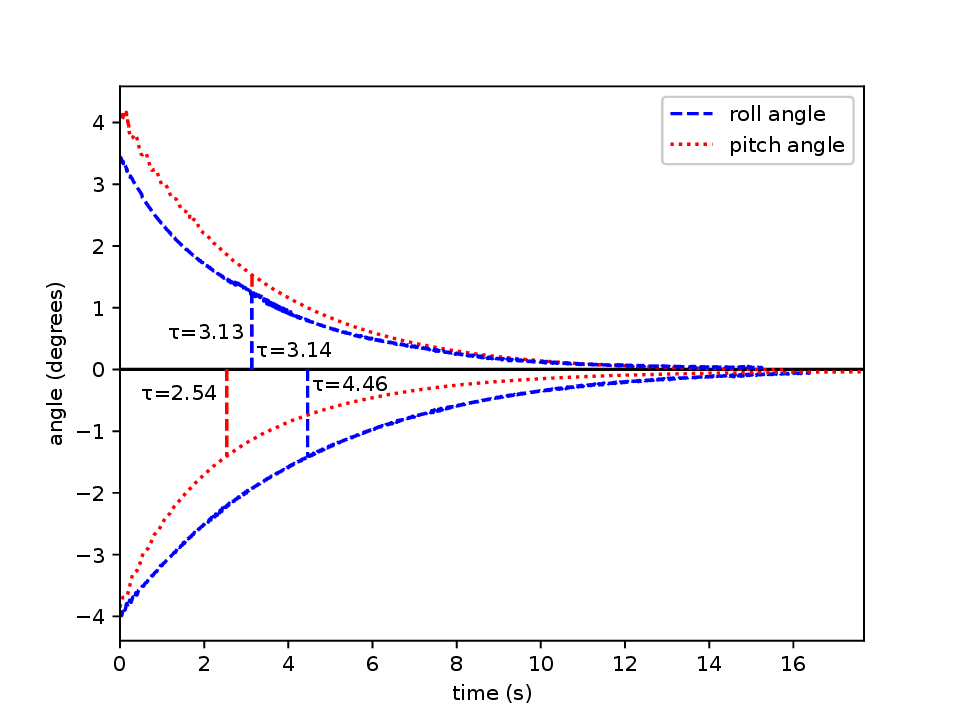}
  \caption{}
  \label{fig:sub1}
\end{subfigure} \\
\begin{subfigure}{\linewidth}
  \centering
  \hspace{-.1\linewidth}
  \includegraphics[width=.8\columnwidth]{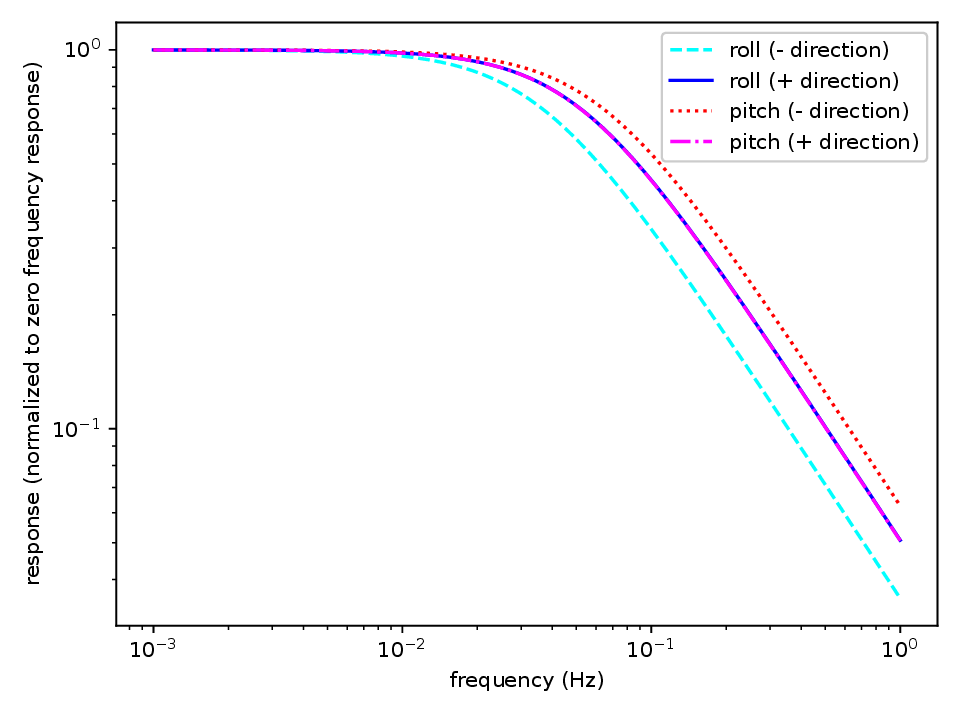}
  \caption{}
  \label{fig:sub2}
\end{subfigure}
\caption{Time constant and frequency response of the flexured-gimbal angle measurement system. (a) Experimental measurements of the flexured-gimbal device time constant. (b) Corresponding frequency response plots for the flexured-gimbal device. The cutoff frequency is approximately 0.03~Hz.}
\label{timeConstants}
\end{figure}
\subsection{Flexured-gimbal Design and Fabrication}
The physical device is shown in Fig.~\ref{deviceLabels}. Damping is added to the system using a carbon fiber rod immersed in a dish of glycerin. This attenuates undesirable resonant oscillations. The gimbal support structure is composed of lightweight laser-cut balsa wood and placed on a precision balance (Mettler-Toledo). The gimbal and FIR are supported to the side of the balance, which does not affect force readings (Section~\ref{sec:force}). They are placed high enough (6.5~cm) to avoid the aerodynamic ground effect. 

The joints of the system are made of a flexible layer of 12~$\mu$m Kapton sandwiched between two 254~$\mu$m fiberglass (FR4) layers. All flexures are oriented vertically and have their primary load in tension to minimize off-axis forces and the chance of buckling. All parts were machined with a diode-pumped 355~nm solid-state laser system (Photomachining, Inc., Pelham, NH). Machining the larger gimbal ring required performing two separate machining steps with the galvo-steered laser, with a precise stage move in between. This was because it is too large to fit within the 50~mm cutting area reachable by the galvo. The two FR4 layers were then bonded with the Kapton layer using Pyralux FR-1500 (Dupont) adhesive sheets. We aligned all layers using tight-fit pins and pressed them under 50~kgf force at 200~C to bond. Once the gimbal is assembled, the ``table'' was attached with CA glue, positioned slightly below the flexure joints, so that when the FIR is mounted on its legs, its COM will be approximately coincident with the the roll and pitch axes of rotation of the gimbal. Our design files are available at https://github.com/aaron-weber-314/FIR-Flexured-Gimbal.

\subsection{Data Collection}

We used Simulink Real-time (Mathworks, Natick, MA) running on a desktop computer to collect data and drive the FIR's piezo actuators. The waveform was produced using analog outputs from a signal acquisition board (PCI-6229, National Instruments, Austin TX USA) at 10~kHz, amplified 50$\times$ using three piezo amplifiers (Trek 2205). We saved orientation data from motion capture at its 240~Hz native frame rate (four Prime 13 cameras, Optitrak, Oregon USA), while force measurements were transmitted at 10~Hz over RS232 from the precision balance (Mettler Toledo, Greifensee, Switzerland). 

\subsection{Sensor Bandwidth}
We estimated the bandwidth of the system by manually moving it by hand to different initial pitch and roll angles and measuring its rate of decay back to its resting (zero) angle using the motion capture system. An FIR was mounted on the device during this process, but it was not active. The time constants were in the range of 3 to 4 seconds, corresponding to  a bandwidth between 0.036~Hz and 0.063~Hz (Fig.~\ref{timeConstants}). This is too slow to measure the effect of a single wingbeat, but sufficient to measure stroke-averaged torques.

\subsection{Robofly Platform}

The UW Robofly platform was introduced in~\citet{ddhingraTrimming}, an enlargement of the design in~\citet{ChukewadTRO}. It weighs 180~mg including motion capture markers and flaps its wings at 180 beats per second. The design of the FIRs used in our experiments has not varied significantly from these works.

\subsection{Experimental Environment}\label{sec:environment}
Experiments with the FIRs were performed in a small motion capture chamber enclosed on all sides but the front to minimize disturbances. In the case of free-flight experiments, the FIR was suspended well above the floor ($> 20$ cm) by a thin tether attached to the top of the chamber to minimize ground effects. Similarly, the base of the gimbal device was elevated to minimize ground effects when taking measurements with an FIR attached

\subsection{Calibration and Resolution}\label{sec:calibration}
\begin{figure}[tbp]
    \includegraphics[width=.9\columnwidth]{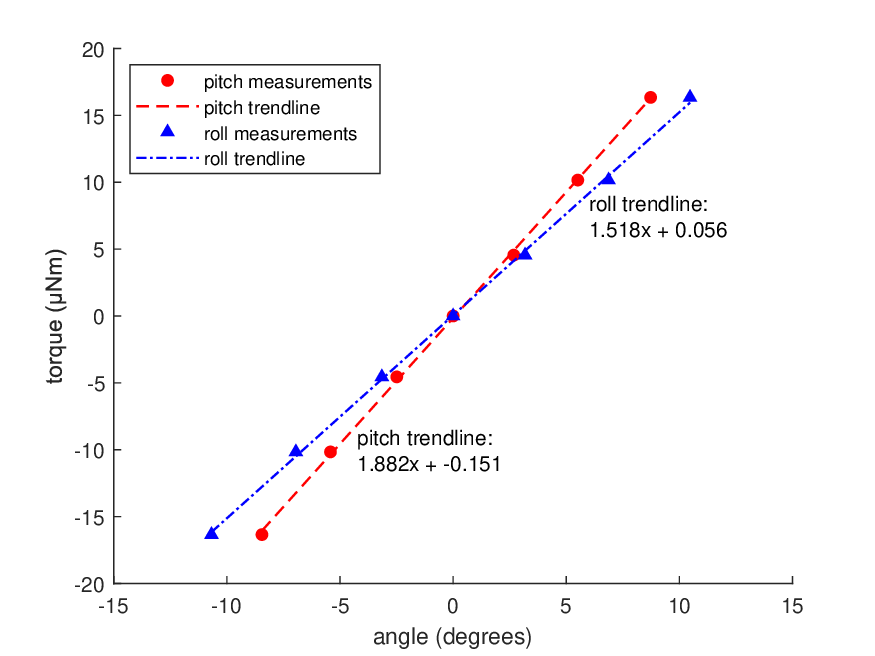}
    \caption{Flexured-gimbal sensor total device stiffness measurements in the roll and pitch axes, with calculated trend lines.}
    \label{calibration_fig}    
\end{figure}

Our desired torque resolution was at least 0.3~$\mu$Nm, which was motivated by the estimated torque uncertainty induced by the thin wire tether that provides power and control signals to the robot~\citep{fuller2014controlling}. Previous work~\citep{Merriaux2017} found the measurement variability of a small motion capture arena to be 0.025~mm. Our motion capture arena is of similar size and uses cameras with half the resolution, so we assume a variability of 0.05~mm. With a distance of 40~mm between the motion capture markers on the device, this corresponds to an angular resolution of 0.14~deg. The stiffness of the flexures alone was insufficient, so to obtain the desired total device stiffness $k_s$ (Eq.~\eqref{eq:sensitivity}) we added a counterweight of mass approximately 50~mg at distance approximately 17~mm below the FIR's COM. To measure $k_s$, we applied known torques by hanging known masses at various known distances from the axis of rotation and measured the resulting static angular deflections. We applied torques in both clockwise and counterclockwise directions. A least-squares linear regression (Fig.~\ref{calibration_fig}) gives a total device stiffness $k_s=1.52~\mu$Nm/rad for roll and $k_s=1.88~\mu$Nm/rad for pitch. With the angular resolution of the motion capture arena, this corresponds to a resolution of 0.27~$\mu$Nm in pitch and 0.22~$\mu$Nm in roll, which is within our design specification.

\subsection{Sensing FIR Torques Using the Flexured-gimbal Sensor}\label{sec:measurement}
\begin{figure}
    \centering
    \includegraphics[width=\linewidth]{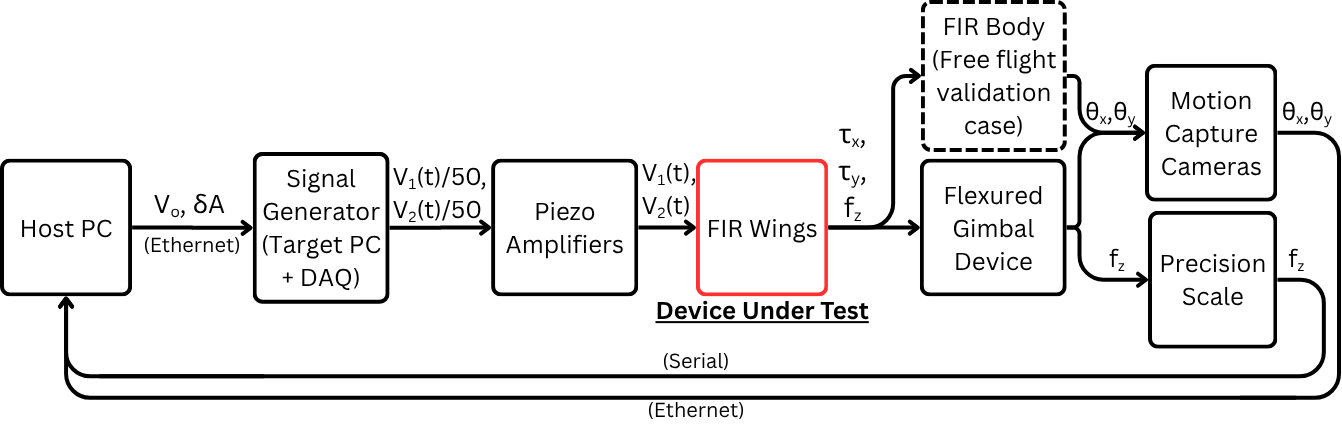}
    \caption{Block diagram of the subsystems and information flow involved in the measurement process.}
    \label{block_diagram}
\end{figure}

Before taking measurements, the FIR is precisely positioned and then bonded to its table as described in Section~\ref{sec:introduction}
and in more detail in~\citet{ddhingraTrimming}, to ensure its COM coincides with the gimbal's two flexure axes. To perform a measurement (Fig. \ref{axisFig}a), a constant command signal sinusoid with a specified $V_o$ and $\delta A$ is applied to the FIR for 3~s. Thrust, pitch angle, and roll angle measurements from the last 0.5~s are taken and averaged, after the transients have died out to ensure only steady-state values are measured. Eq.~\eqref{eq:sensitivity} and the calibration derived in Section~\ref{sec:calibration} are used together to estimate pitch and roll torques. The connections and information flow between subsystems in the measurement process is illustrated in Fig.~\ref{block_diagram}.

\subsection{Free-flight Trim Torque Validation}\label{sec:methods_validation}
``Ground truth" torque measurements were obtained by applying either zero load torque, or applying a known static free-flight torque by attaching a small offset weights to a rod at a known distance from the robot's center of mass (COM) (Fig.~\ref{axisFig}b) and trimming the robot in free flight. To do so, we found trim voltages $\delta A$ and $V_o$ by iteratively changing voltage command values until the FIR took off nearly vertically. This manual trimming operation is repeatable to within about 1~V. This trimming process is in general required to compensate for undesired roll and pitch bias torques that result from manufacturing variability. We favor this means of measuring torques in free flight over the alternative of estimating angular acceleration for two reasons. First, it does not require knowing the moment of inertia of the vehicle. And second, it does not require estimating angular acceleration, which is subject to considerable noise due to the need to take two numerical derivatives of the orientation measurements from motion capture data.

\section{Results}

We performed measurements and sensor validation using two different Roboflies. The first, the ``mapping fly," was used to take the map of input-output commands across a range of commands as described below. This entailed extended operation time, and this fly failed shortly afterward due to a crack in one of its piezoelectric actuators, as observed by a substantial reduction in thrust. We therefore constructed a second Robofly, the ``validation fly," that we used to validate our flexured-gimbal sensor readings against static free flight torque measurements. 

\subsection{Preliminary Free-flight Validation Torque Measurement }\label{sec:validation}

Before attaching the two flies to the torque sensor gimbal, we performed preliminary free-flight manual trimming procedures on both  (Section~\ref{sec:methods_validation}) under condition of zero applied torque to help validate our results. The mapping fly's trim values were found to be $\delta A = 17$~V for roll and $V_o= -3$~V for pitch. The roll trim value is large but not uncommon. Similarly, trim values for the validation fly were $\delta A = -10$~V for roll and $V_o= 7.5$~V. For the validation fly, we performed additional free-flight validation tests under conditions of a known applied torque before mounting it to the torque sensor.  With a 31.8~mg mass mounted on a rod extending 4~mm from the COM along the pitch axis, corresponding to a 1.25~$\mu$Nm roll torque, the required roll trim voltage changed to $\delta A = -8.5$~V. Similarly, a 25~mg mass was added to a rod extending 4~mm from the COM along the roll axis, corresponding to a $-0.98$~$\mu$Nm pitch torque. The required pitch trim voltage changed to $V_o = 5$~V. After being bonded to the force/torque sensor for measurements, the validation fly was detached and trimmed one final time, giving a slightly-changed $\delta V = -11.5$~V for roll and $V_o = 5$~V for pitch. The change in trim was presumably attributable to wear in the FIR's flapping mechanism. 

\subsection{Torque Measurements on the Gimbal}
After trimming in free flight, both FIRs were mounted on the flexured-gimbal torque sensor. The mapping fly was actuated across a grid of different pitch and roll commands while pitch torque, roll torque, and thrust were measured  as described in Section~\ref{sec:measurement}. All combinations of roll and pitch voltages were tested with the exception of the very extreme corners, where the simultaneous application of maximum pitch and roll commands would have exceeded the 250~V bias rail being supplied to the actuators. The baseline peak-to-peak amplitudes (equal to $2A$ in Eqs.~(\ref{v1_eq}),~(\ref{v2_eq})) of 192~V for the mapping fly and 168~V for the validation fly were the values at which the vehicle just barely lifted off.

\begin{figure}[tbp]
    \begin{subfigure}{.9\columnwidth}
        \includegraphics[width=\columnwidth]{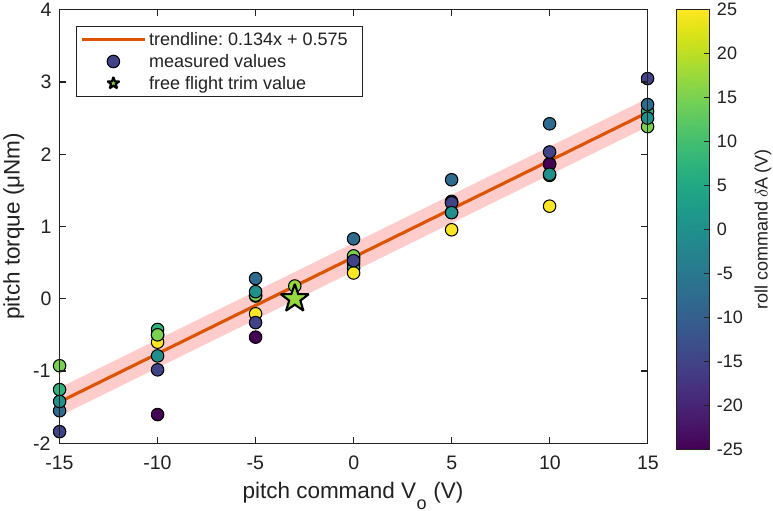} 
        \caption{ }
        \label{pitch1}
    \end{subfigure}
    \begin{subfigure}{.9\columnwidth}
        \includegraphics[width=\columnwidth]{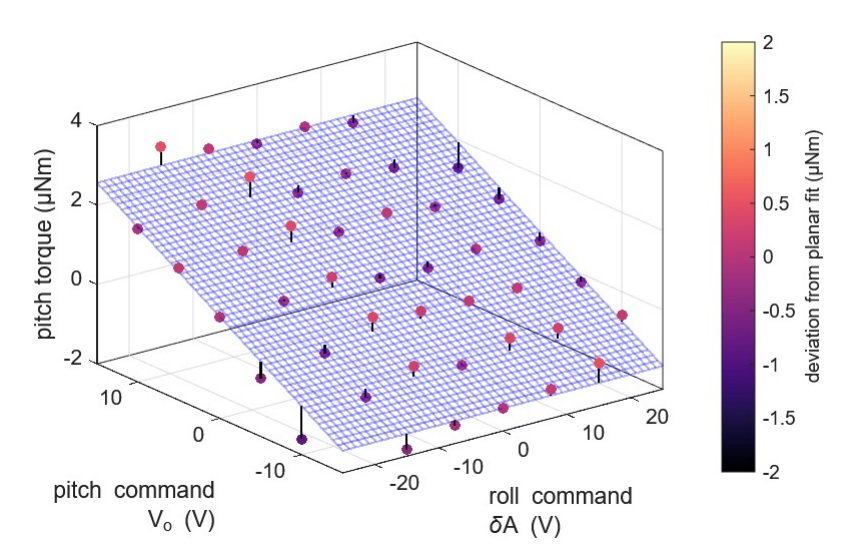}
        \caption{ }
        \label{pitch3d}
    \end{subfigure}
    \caption{Pitch torque is not significantly impacted by changes in roll command. (a) Pitch torque produced by the FIR as measured by the gimbal sensor, as a function of both pitch command (offset voltage $V_o$) and roll command (amplitude difference $\delta A$). A color map shows the strength of the roll command at each data point. The free flight zero-pitch-torque trim value (denoted by a star) is close to the trend line, consistent with the torque measurements. Shaded region shows $\pm$ one standard deviation ($\sigma_{pitch}$) away from the trend line. (b) A 3D representation of the same data, with a 2D regression planar fit. Data points are colored to indicate the size of the deviation from the planar fit. }
    \label{pitchCombo}
\end{figure}

\begin{figure}[tbp]
    \begin{subfigure}{.9\columnwidth}
        \includegraphics[width=\columnwidth]{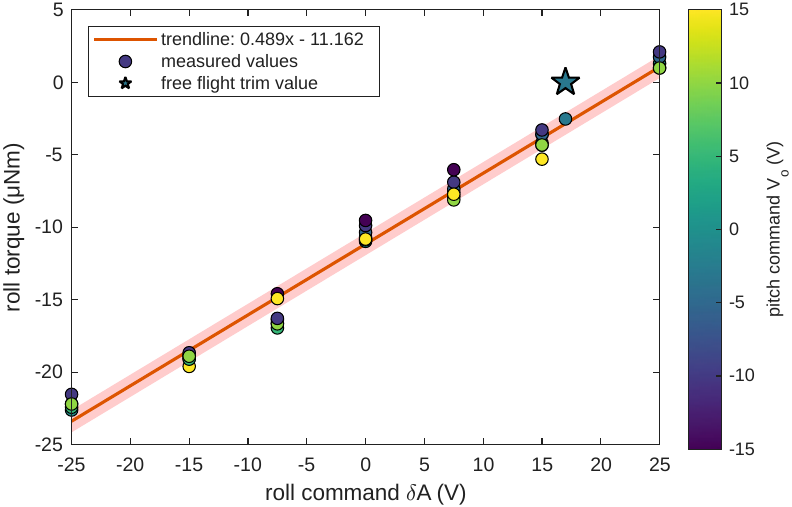} 
        \caption{ }
        \label{roll1}
    \end{subfigure}
    \begin{subfigure}{.9\columnwidth}
        \includegraphics[width=\columnwidth]{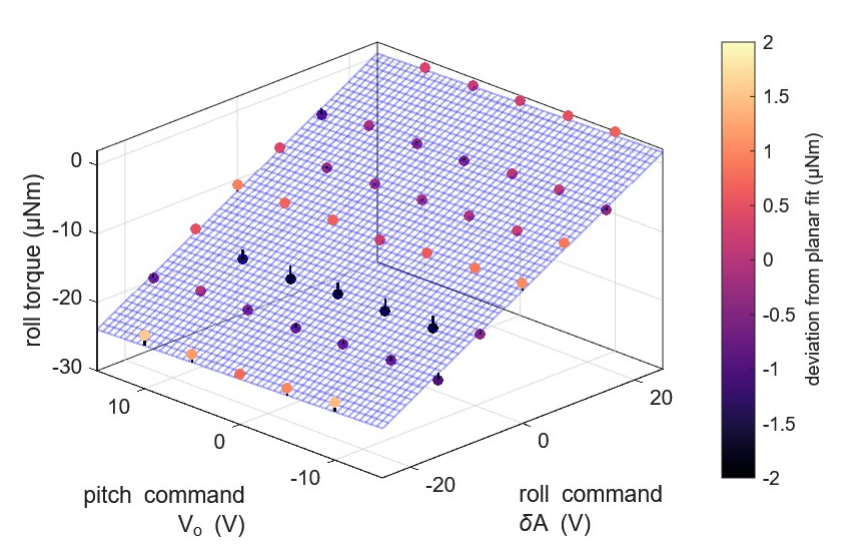}
        \caption{ }
        \label{roll3d}
    \end{subfigure}
    \caption{Roll torque is not significantly impacted by changes in pitch command. (a) Roll torque produced by the FIR as measured by the gimbal sensor, as a function of both pitch command (offset voltage $V_o$) and roll command (amplitude difference $\delta A$). A color map shows the strength of the pitch command at each data point. The free flight zero-roll-torque trim value (denoted by a star) is close to the trend line, consistent with the torque measurements. Shaded region shows $\pm$ one standard deviation ($\sigma_{roll}$) away from the trend line. (b) A 3D representation of the same data, with a 2D regression planar fit. Data points are colored to indicate the size of the deviation from the planar fit. }
    \label{rollCombo}
\end{figure}

Figs.~\ref{pitch1} and \ref{roll1} show how pitch and roll commands affect the pitch and roll torque applied by the mapping fly. A least-squares linear regression of pitch torque vs. pitch command $V_o$ has a slope of 0.13~$\mu$Nm/V and an $R^2$ = 0.95 coefficient of determination, indicating strong correlation. Similarly, roll torque vs. roll command $\delta A$ has a slope of 0.49~$\mu$Nm/V and $R^2$ = 0.98. To compare behavior inside and outside extreme control voltage ranges, we computed the standard deviations from the trend line of the nine commands with the smallest magnitude, giving $\sigma_{roll}$ = 0.77 and $\sigma_{pitch}$ = 0.20  (shown as a shaded area in the figures). At the extreme control points ($\pm$25~V roll, $\pm$15~V pitch), the standard deviation from the trend line was 3.46 in roll and 0.32 in pitch, showing that there is more variability in output torques in that range. It should be noted that the measured roll torque outputs in Fig.~\ref{roll1} are almost entirely negative, which is not an issue with the measurement process but rather a reflection of the manufacturing variability inherent in these FIRs that makes trimming essential for control. The mapping fly used for these measurements was heavily biased towards the negative roll direction and the torques it was able to actuate reflect that, but the mapping itself is still accurate as those biases are caused by the FIR itself and not the device.

Measurements on a second Robofly, the validation fly, show that the mapping from input to output on the flexured-gimbal sensor were slightly different but comparable, at 0.11~$\mu$Nm/V (compared to 0.13) for pitch and 0.38~$\mu$Nm/V (compared to 0.49) for roll (Figs.~\ref{rollVal} and~\ref{pitchVal}). Additionally, our torque mapping results are similar to measurements on a third Robofly in \citep{dhingra2025modeling}, which were 0.11~$\mu$Nm/V for pitch and 0.48~$\mu$Nm/V for roll.

\begin{figure}[tbp]
    \begin{subfigure}{.9\columnwidth}
        \includegraphics[width=\columnwidth]{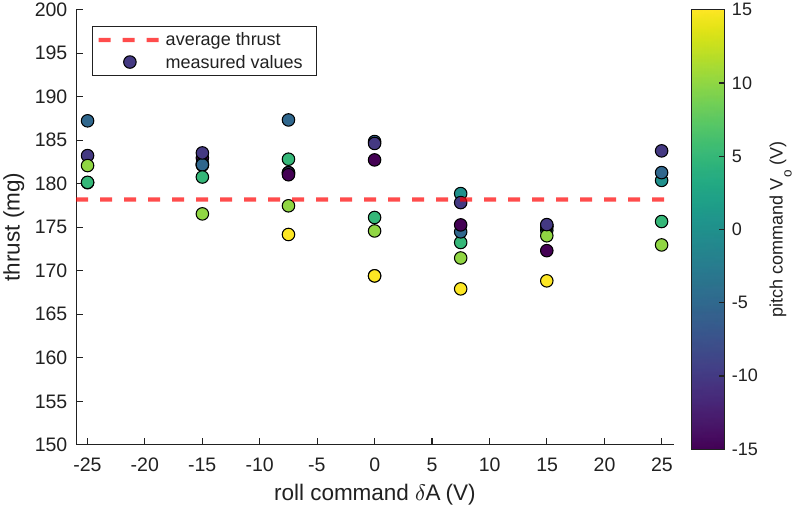}
        \caption{ }
    \end{subfigure}
    \begin{subfigure}{.9\columnwidth}
        \includegraphics[width=\columnwidth]{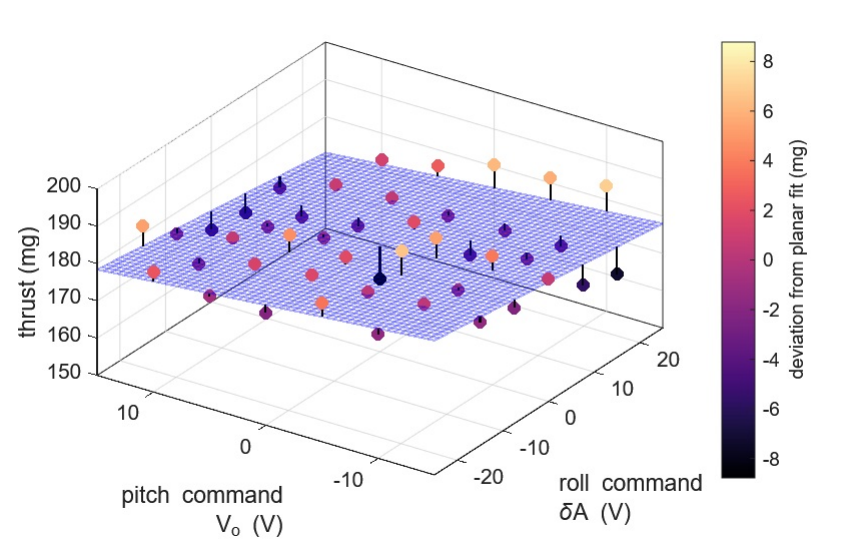}
        \caption{ }
    \end{subfigure}
    \caption{Thrust force is not significantly impacted by roll and pitch commands, though there is a discernible trend of reducing thrust as pitch command increases. (a) Thrust force as measured by precision balance as a function of pitch and roll command. A colormap shows the deviation from the mean thrust value. (b) A 3D representation of the same data, with a 2D regression planar fit. Data points are colored to indicate the  size of the deviation from the planar fit.}
    \label{thrustFig}
\end{figure}

\subsubsection{Pitch and Roll Torque Coupling}
To more thoroughly characterize the extent to which roll commands have an effect on pitch torque and vice versa, Figs.~\ref{pitch3d} and \ref{roll3d} show a representation of a 2D linear regression (``planar fit''). Cross-correlation coefficients were $-$0.001 between roll commands and pitch torques
and $-$0.058 between pitch commands and roll torques, indicating negligible cross-axis coupling. A visual inspection shows that roll commands at either extreme appear to cause pitch torque to sometimes be more negative than the command (i.e., pitch more backward), if there is also simultaneously a strong pitch command. A possible explanation that flexures in the wing's transmission may be reaching the limit of the linear regime. No such coupling is observed for roll torques, however. 

\subsubsection{Relation between thrust and control voltages}
As seen in Fig.~\ref{thrustFig}, there is some variation in the thrust with the varying control voltages, but the maximum deviation from the mean thrust is small (5.8\%). There is only a weak trend connecting roll and pitch voltages to thrust (a slope of $-$0.26~mg/V in pitch and $-$0.16~mg/V in roll).

\subsection{Post-Sensor Validation}
\begin{figure}[tbp]
    \includegraphics[width=.9\columnwidth]{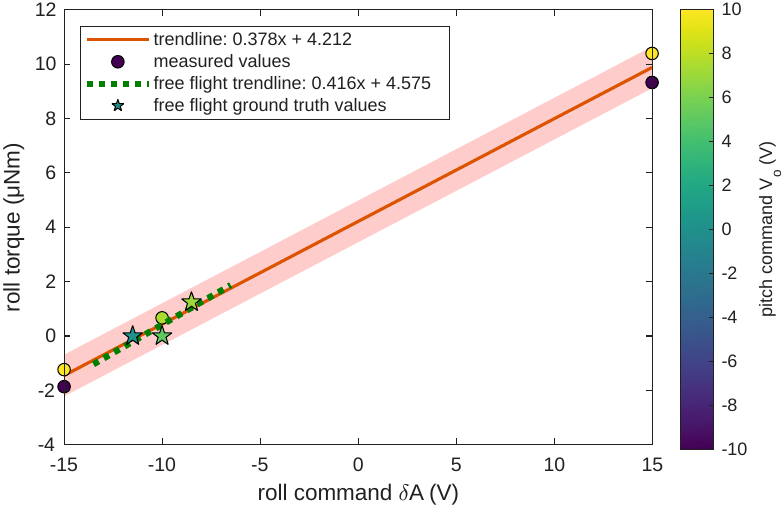}
    \caption{Validation results on a second FIR show a similar voltage-to-torque roll torque slope and decoupled torque output both on the sensor and in free flight. Shaded region shows $\pm$ one standard deviation from the original mapping fly ($\sigma_{roll}$) away from the trend line.}
    \label{rollVal}
\end{figure}

\begin{figure}[tbp]
    \includegraphics[width=.9\columnwidth]{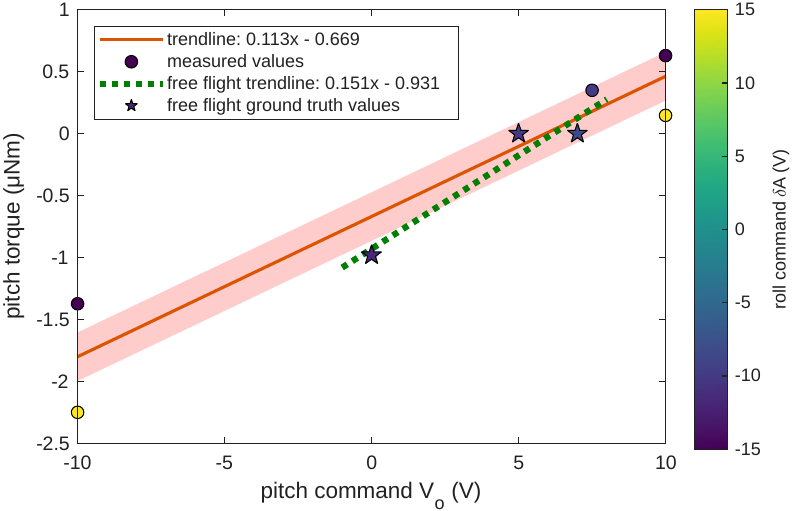}
    \caption{Validation results on a second FIR show a similar voltage-to-torque pitch torque slope and decoupled torque output both on the sensor and in free flight. Shaded region shows $\pm$ one standard deviation from the original mapping fly ($\sigma_{pitch}$) away from the trend line.}
    \label{pitchVal}
\end{figure}

\begin{figure}[tbp]
\centering
\begin{subfigure}{.5\linewidth}
  \centering
  \includegraphics[width=.95\linewidth]{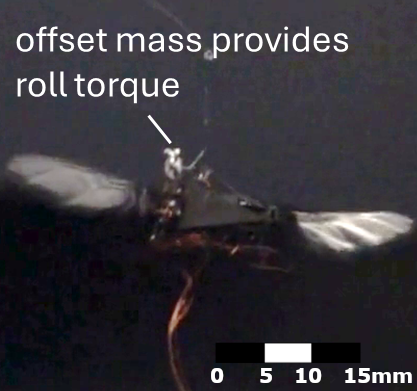}
  \caption{}
  \label{fig:sub1}
\end{subfigure}%
\begin{subfigure}{.5\linewidth}
  \centering
  \includegraphics[width=.95\linewidth]{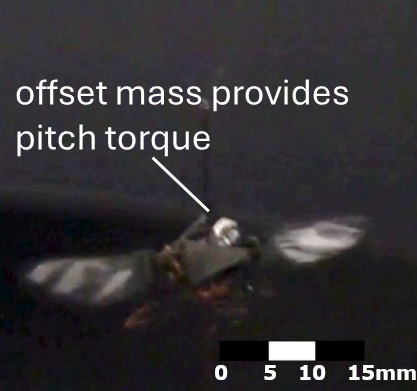}
  \caption{}
  \label{fig:sub2}
\end{subfigure}
\caption{Validation of torque mapping in free flight using known offset torques. Control values to oppose these torques and achieve a straight takeoff are compared to the control mapping results. (a) A mass offset from the center rod in the roll axis provides known roll torque. (b) A mass offset from the center rod in the pitch axis provides known pitch torque. Videos available at https://youtu.be/COBHHIgYGRU}
\label{valPhotos}
\end{figure}

We took five torque measurements of the validation fly's torques using the flexured-gimbal sensor. Following these measurements, the validation fly was then removed from the gimbal and then trimmed in free flight again, as described in Section~\ref{sec:methods_validation}. 
The trim values changed a small amount, presumably due to wear, resulting in two different trim values in free flight, one from before the measurement and one from after (Figs.~\ref{rollVal} and \ref{pitchVal}). Finally, the FIR was trimmed under conditions in which an offset mass was added to apply a known static torque load (Section~\ref{sec:methods_validation}) about either the pitch or roll axis, shown in Fig.~\ref{valPhotos}. Figs.~\ref{rollVal} and \ref{pitchVal} show this torque estimate as a star superimposed on the gimbal sensor's measurements. The free flight values' deviation from the validation trend lines largely fall within the $\sigma_{roll}$ and $\sigma_{pitch}$ values from the mapping fly, validating those measurements. The single error outside that range, occurring at 0~V pitch, is about equal to the typical torque disturbance of the wire tether measured in~\citet{fuller2014controlling}  of 0.3~$\mu$Nm. 

\section{Conclusion and Future Work}

This paper introduces a device design and measurement process that is designed to conveniently measure the mapping from control commands to roll and pitch torques produced in sub-gram flapping-wing flying insect robots (FIRs). This device is the first sensor capable of taking two-axis torque measurements simultaneously that is sensitive enough to be used with FIRs. While the device shown in this paper was designed for taking measurements with the UW Robofly, the flexures and gimbal can readily be scaled for use with any flapping-wing robot. We have published the design files for the device should anyone wish to apply it to other FIRs at {https://github.com/aaron-weber-314/FIR-Flexured-Gimbal}. This system is an improvement over an earlier system in~\citet{ddhingraTrimming} that was only capable of finding compensatory trim values rather than measuring torques directly. Like that system, the system here is constructed entirely using parts that are likely available in a lab or factory creating FIRs. Its gimbal and flexures are machined using the same laser system used to construct the robot itself. Readout is performed using a motion capture system that is standard equipment in many robotics settings. As a consequence of this design, however, only low-frequency measurements with a maximum bandwidth of approximately 0.3 Hz are possible.

Further improvements to the device to decrease cost and increase ease of use could be made by substituting the motion capture system with a low-cost micro-electromechanical (MEMS) accelerometer as an inclinometer. It is expected that most if not all fully-autonomous FIRs will already have an accelerometer as an integral part of their inertial navigation system. Our preliminary work indicated that vibrations due flapping wings confound the measurements, but we anticipate this could be mitigated by positioning the accelerometer on the damping rod near the glycerin. 
While motion capture and MEMS inclinometers can in principle provide sufficient bandwidth to measure transient torques on a wingstroke-by-wingstroke basis, the low stiffness $k_s$ inherent to our system is what limits its bandwidth. For higher bandwidth, a more costly capacitive sensor, which has a higher stiffness, may be needed~\citep{finio2011torques}. However, obtaining an equivalent cross-axis torque map would require double the flapping time because such sensors have only demonstrated torque sensing around a single axis. Measuring the second axis would require removing the FIR and re-orienting it on the sensor. 

Our results show that the roll and pitch torques of the flying insect robot platform we tested, the 180~mg piezo-actuated UW Robofly, are essentially decoupled and therefore can be actuated independently. These findings were validated on two different Robofly devices. We therefore have validated a common control system approach in which longitudinal and lateral dynamics are controlled independently. We anticipate this new information can additionally be used to better model the dynamics of flapping-wing robots and control their movements more effectively, especially when undergoing aggressive (high-torque) maneuvers.

\begin{ack}
We thank Santosh Devasia for insightful discussions during the writing of this manuscript. This work was partly supported by the NSF award Nos. 2319060, 2054850, and 2235207.
\end{ack}

\bibliography{ifacconf}             
\end{document}